\documentclass[aps,preprint,pre,superscriptaddress,runinaddress,showpacs]{revtex4}
\usepackage{amssymb}
\usepackage{amsmath}
\usepackage{graphicx}
\usepackage{bm}

\begin{document}

\title{Theoretical Description of Coulomb Balls - Fluid Phase}
\author{J. Wrighton}
\author{J. W. Dufty}
\affiliation{Department of Physics, University of Florida, Gainesville, FL 32611}
\author{H. K\"ahlert}
\author{M. Bonitz}
\affiliation{Institut f\"{u}r Theoretische Physik und Astrophysik, Christian-Albrechts
Universit\"{a}t zu Kiel, 24098 Kiel, Germany}
\date{\today }

\begin{abstract}
A theoretical description for the radial density profile of a finite number
of identical charged particles confined in a harmonic trap is developed for
application over a wide range of Coulomb coupling (or, equivalently,
temperatures) and particle numbers. A simple mean field approximation
neglecting correlations yields a density profile which is monotonically
decreasing with radius for all temperatures, in contrast to molecular
dynamics simulations and experiments showing shell structure at lower
temperatures. A more complete theoretical description including charge
correlations is developed here by an extension of the hypernetted chain
approximation, developed for bulk fluids, to the confined charges. The
results reproduce all of the qualitative features observed in molecular
dynamics simulations and experiments. These predictions are then tested
quantitatively by comparison with new benchmark Monte Carlo simulations.
Quantitative accuracy of the theory is obtained for the selected conditions
by correcting the hypernetted chain approximation with a representation for
the associated bridge functions.
\end{abstract}

\maketitle


\section{Introduction}

Strong correlation effects in ensembles of charged particles are
important in many fields of physics, including plasmas, the electron
gas in solids, and electron-hole plasmas, e.g.
\cite{dubin99,bonitz2008pop} and references therein. In recent years
charged particles spatially confined in trapping potentials have
attracted considerable interest. Examples are ultracold ions
\cite{Wineland1987,drewsen1998}, dusty plasmas
\cite{Arp2004,bonitz2006}, and electrons in quantum dots
\cite{Alex2001}. Related studies of expanding neutral plasmas  have
been reported as well \cite{Pohl2004,Killian}. Recently, the ground
state structure of Coulomb crystals formed by classical charges in a
harmonic trap has been studied in detail via molecular dynamics (MD)
simulation for Coulomb interaction in \cite{Ludwig2005,Arp2005} and
for Yukawa interaction in \cite{bonitz2006,Baumgartner2008}. The
same system is studied here for its fluid phase at finite
temperatures. For most experimental situations, e.g. in ultracold
ions and dusty plasmas, finite temperature effects are expected. The
objective here is to provide a theoretical description for the
structure of this system over the full range of temperatures for the
fluid phase.

Among the primary properties of interest is the radial density profile of
the particles in the trap. A theoretical description of the ground state has
been obtained via an energy minimization for both Coulomb and Yukawa systems
without correlations \cite{Christian2006}, and with correlations in a local
density approximation \cite{Christian2007}. Good agreement with simulations
for the spatial average density profile was obtained in this way, but a more
complete representation of correlations is required for the detailed spatial
modulation (shells) expected at all except the highest temperatures. This is
confirmed by the theoretical analysis developed here, following \cite{sccs08}%
. This theory is parameterized by the number of particles in the trap, $N$,
and the plasma coupling constant, $\Gamma $, measuring the strength of the
Coulomb energy for a pair of charges in the trap relative to the kinetic
energy (defined below). Alternatively, the average kinetic energy relative
to trap energy or to Coulomb energy can be used to define a dimensionless
temperature $T^{\ast }\equiv 1/\Gamma $. The range of values explored is $%
1<N\leq 300$ and $0<\Gamma \leq 40$ (or $0.025 \leq T^{\ast } < \infty $).

The primary results described here are: 1) an extension of the HNC
approximation developed for bulk fluids to systems with localized densities,
within the context of density functional theory; 2) a confirmation that mean
field theory (no correlations) predicts a structureless, monotonically
decreasing radial profile approaching a step function at $T^{\ast }=0$; 3) a
demonstration that HNC correlations lead to a shell structure (local radial
peaks) for strong coupling $\Gamma \geq 10$ $\left( T^{\ast }\leq 0.1\right)
$; 4) new Monte Carlo simulations to test this theory; and 5) a proposal for
corrections to the HNC that leads to predictions in good quantitative
agreement with the Monte Carlo simulations over the range of $N,\Gamma $
tested.

The qualitative features of the shell structure arising from correlations
are: the number of shells increases with $N$, with new shells appearing at
special values of $N$; a sharpening of the shells (more narrow, higher
amplitude) with increasing $\Gamma $ (decreasing $T^{\ast }$); and shell
populations that grow monotonically with $N$ but which are almost
independent of $\Gamma $ ($T^{\ast }$) (as observed in previous MD
simulations \cite{Golubnychiy2006,Baumgartner2007}). The HNC approximation
provides agreement on all qualitative features of the density profile, e.g.
the location of the shells and number of particles within the shells is
well-described at all coupling values. However, discrepancies with the Monte
Carlo simulations of the order of $20-40\%$ for the widths and amplitudes of
the shell structure are found. These discrepancies are removed by including
a representation of the "bridge functions" neglected in the HNC
approximation, referred to in the following as the adjusted HNC (AHNC). The
AHNC theory has the simplicity of the HNC, depending only on well-known
correlations of the bulk one component plasma, and shows excellent agreement
with simulations.

This paper is organized as follows. First the Hamiltonian for the system is
defined and the density profile is formulated in the Canonical ensemble. The
relevant energy and length scales are introduced for a dimensionless
representation in terms of $N$ and $\Gamma $ (or $T^{\ast }$). This is the
form most appropriate for Monte Carlo simulation. In Section \ref{sec3}, the
corresponding formulation is given in the Grand Canonical ensemble to allow
exploitation of density functional theory. A formally exact representation
for the density profile is obtained in Appendix A in terms of correlations
in a bulk one component plasma (OCP), together with a corresponding exact
equation for the OCP. The HNC approximation is defined in Section \ref{sec3}
as the neglect of the "bridge functions" in these two sets of equations. A
further neglect of all correlations, the mean field approximation, is
explored first over the whole temperature range for reference in Section \ref%
{sec4}. The effects of correlations within the HNC approximation are
illustrated in Section \ref{sec5}. Selected results are compared with those
from Monte Carlo simulations in Section \ref{sec6}, where the AHNC is
defined and shown to give the necessary corrections to HNC required for
quantitative agreement. Finally the results are summarized and discussed in
the last Section.

In closing this Introduction, it is worth emphasizing that the interesting
radial structures studied occur for strong plasma coupling, $\Gamma \gtrsim
10$. Thus the system presents a good example of a strongly coupled Coulomb
system. For very large $\Gamma $ the shells become effectively systems of
particles uniformly distributed on the surfaces of spheres of specific
radii. At some point the rotational symmetry on these spherical surfaces is
broken and these particles form spherical Wigner crystals \cite{bonitz2006}.
An adequate description of correlations within density functional theory
should provide both the fluid phase considered here and the freezing
transition. The crystal phase is more subtle than in a bulk plasma since it
entails packing on a non-Euclidean surface, with necessary disclinations
depending on the value of $N$ considered (related to Thomson's problem \cite%
{Bowick}).

The determination of the non-uniform density for charges confined in a trap
is similar to the complementary problem of determining the enhanced density
in an OCP with an impurity ion of the opposite charge. In the latter case
the density profile of the bound states corresponds to that for the charges
in the trap. The formalism here extends to this case in a natural way \cite%
{sccs08}. Both the structure and dynamics for this problem have been studied
in some detail recently \cite{Talin}, and will not be considered further
here.

\section{Monte Carlo Simulation}

\label{sec2}

A system of $N$ identical charges in a spherical container of radius $R$ is
considered. An attractive central force is applied at the origin. The
Hamiltonian is
\begin{equation}
H=\sum_{i=1}^{N}\left( \frac{1}{2}mv_{i}^{2}+V_{0}\left( r_{i}\right)
\right) +\frac{1}{2}\sum_{i\neq j=1}^{N}V(r_{ij})+V_{W}.  \label{2.1}
\end{equation}%
Here $\mathbf{r}_{i}$ and $\mathbf{v}_{i}$ are the position and velocity of
charge $i$. The repulsive interaction potential between charges $i $ and $j $
is denoted by $V(r_{ij})$ where $r_{ij}\equiv \left\vert \mathbf{r}_{i}-%
\mathbf{r}_{j}\right\vert $. The single particle potential $V_{0}\left(
r_{i}\right) $ represents the central "confinement" potential (referred to
as the "trap" below) and $V_{W}$ is the wall potential (zero inside the
container and infinite otherwise). The focus here is on the average density
of charges $n(r)$ for the equilibrium state, with spherical symmetry in the
fluid phase. In the classical Canonical ensemble this is defined by%
\begin{equation}
n(r_{1})=N \frac{\int d\mathbf{r}_{2}..d\mathbf{r}_{N}e^{-\beta V(\mathbf{r}%
_{1},..,\mathbf{r}_{N})}}{\int d\mathbf{r}_{1}..d\mathbf{r}_{N}e^{-\beta V(%
\mathbf{r}_{1},..,\mathbf{r}_{N})}}.  \label{2.2}
\end{equation}%
Attention will be focused on a harmonic trap and Coulomb interactions among
charges. The total potential energy in dimensionless form is given by%
\begin{equation}
V^{\ast }(\mathbf{r}_{1}^{\ast },..,\mathbf{r}_{N}^{\ast })\equiv \beta V(%
\mathbf{r}_{1},..,\mathbf{r}_{N})=\Gamma \left[ \frac{m\omega ^{2}r_{0}^{3}}{%
2q^{2}}\sum_{i=1}^{N}r_{i}^{\ast 2}+\frac{1}{2}\sum_{i\neq j=1}^{N}\frac{1}{%
\left\vert \mathbf{r}_{i}^{\ast }-\mathbf{r}_{j}^{\ast }\right\vert }\right]
+V_{W}^{\ast }.  \label{2.4}
\end{equation}%
Here, $\mathbf{r}_{i}^{\ast }=\mathbf{r}_{i}/r_{0}$, and $\Gamma $ is the
Coulomb coupling constant $\Gamma =\beta q^{2}/r_{0}$. The usual choice for
the length scale $r_{0}$ is the ion sphere radius, or mean distance between
charges, given by $4\pi r_{0}^{3}\overline{n}/3=1$, where $\overline{n}$ is
the spatially averaged density for the confined particles. This uniform
density can be characterized by the volume of a sphere inside of which the
particles are localized. This localization can be due to either the walls or
the trap alone, depending on conditions. Confinement by the trap is
determined by the condition that the net force on the outermost charge at
the surface of this sphere should be zero%
\begin{equation}
\frac{Nq^{2}}{R_{0}^{2}}=m\omega ^{2}R_{0},\hspace{0.25in}\Rightarrow
R_{0}^{3}=N\frac{q^{2}}{m\omega ^{2}}.  \label{2.5a}
\end{equation}%
This applies only if $R_{0}<R$; otherwise the particles are confined by the
wall. In the case of interest here, $R_{0}<R$ and the average density is
therefore
\begin{equation}
\overline{n}\equiv \frac{3}{4\pi R_{0}^{3}}\int d\mathbf{r}n(r)=\frac{%
3m\omega ^{2}}{4\pi q^{2}},\hspace{0.25in}r_{0}^{3}=\frac{q^{2}}{m\omega ^{2}%
}.  \label{2.6}
\end{equation}%
Note that the average density is independent of $\Gamma $ and $N$, although
the profile $n(r)$ depends strongly on both.

The dimensionless potential (\ref{2.4})\ is now
\begin{equation}
V^{\ast }(\mathbf{r}_{1}^{\ast },..,\mathbf{r}_{N}^{\ast })=\Gamma \left[
\frac{1}{2}\sum_{i=1}^{N}r_{i}^{\ast 2}+\frac{1}{2}\sum_{i\neq j=1}^{N}\frac{%
1}{\left\vert \mathbf{r}_{i}^{\ast }-\mathbf{r}_{j}^{\ast }\right\vert }%
\right] ,  \label{2.7}
\end{equation}%
and the dimensionless number density is%
\begin{equation}
n^{\ast }(r_{1}^{\ast })\equiv n(r_{1})r_{0}^{3}=N\frac{\int d\mathbf{r}%
_{2}^{\ast }..d\mathbf{r}_{N}^{\ast }e^{-V^{\ast }(\mathbf{r}_{1}^{\ast },..,%
\mathbf{r}_{N}^{\ast })}}{\int d\mathbf{r}_{1}^{\ast }..d\mathbf{r}%
_{N}^{\ast }e^{-V^*(\mathbf{r}_{1}^{\ast },..,\mathbf{r}_{N}^{\ast })}}.
\label{2.8}
\end{equation}%
The parameters characterizing the profile in these units are $\Gamma $ and $%
N $. The coupling constant $\Gamma $ can be interpreted as the inverse
temperature in units of the Coulomb energy $q^{2}/r_{0}$%
\begin{equation}
T^{\ast }\equiv \frac{k_{B}T}{m\omega ^{2}r_{0}^{2}}=k_{B}T\frac{r_{0}}{q^{2}%
}=\frac{1}{\Gamma }.  \label{2.9}
\end{equation}%
The dimensionless density profile $n^{\ast }(r^{\ast })$ can be calculated
numerically by direct Monte Carlo integration of (\ref{2.8}) for given $%
\Gamma $ and $N$.

\section{Theory and Approximations}

\label{sec3}

For the theoretical analysis of the next sections, it is more convenient to
work in the grand Canonical ensemble

\begin{equation}
n(r_{1})=\Xi ^{-1}\sum_{N=0}^{\infty }\frac{N}{N!}\left( \frac{zr_{0}^{3}}{%
\lambda ^{3}}\right) ^{N}\int d\mathbf{r}_{2}^{\ast }..d\mathbf{r}_{N}^{\ast
}e^{-V^{\ast }(\mathbf{r}_{1}^{\ast },..,\mathbf{r}_{N}^{\ast })},
\label{3.1}
\end{equation}%
with the normalization constant%
\begin{equation}
\Xi =\sum_{N=0}^{\infty }\frac{1}{N!}\left( \frac{zr_{0}^{3}}{\lambda ^{3}}%
\right) ^{N}\int d\mathbf{r}_{1}^{\ast }..d\mathbf{r}_{N}^{\ast }e^{-V^{\ast
}(\mathbf{r}_{1}^{\ast },..,\mathbf{r}_{N}^{\ast })}.  \label{3.1a}
\end{equation}%
Here $z=e^{\beta \mu }$, $\mu $ is the chemical potential, and $\lambda
=\left( h^{2}\beta /2\pi m\right) ^{1/2}$ is the thermal wavelength. This
choice for the ensemble allows application of methods from density
functional theory \cite{Evans,Lutsko}. The relevant exact equations are
summarized in Appendix \ref{apA}. In particular, the density profile $n(r)$
is related to the excess free energy $F_{ex}\left( \beta \mid n\right) $
expressed as a functional of that density%
\begin{equation}
\ln \frac{n(r)\lambda ^{3}}{z}=-\beta V_{0}\left( r\right) -\frac{%
\delta \beta F_{ex}\left( \beta \mid n\right) }{\delta n\left( r\right) }.
\label{3.2}
\end{equation}%
The excess free energy is a universal functional of the density such that
each choice for the density corresponds to a different applied external
potential. The case of interest here is the non-uniform density resulting
from the harmonic trap, $\beta V_{0}\left( r\right) \rightarrow $ $%
\Gamma r^{\ast 2}/2$. A different case is a uniform density, which results
from an applied potential of a uniform neutralizing charge density. This
latter case is the familiar one component plasma (OCP) \cite{Hansen}. Both
are described by (\ref{3.2}) with the same excess free energy functional;
only the external potential is different.

In Appendix \ref{apA} two exact representations for $F_{ex}\left( \beta \mid
n\right) $ are given in terms of a uniform reference density. In the first case $V_{0}\left( r_{\alpha }\right) $ is chosen to be the
Coulomb interaction for an ion placed at the origin, plus the uniform
neutralizing background charge. The density in (\ref{3.2}) then becomes the
pair distribution function $g_{OCP}^{\ast }(r^{\ast })$ in the OCP. The
appropriate reference density is its uniform value far from the ion, for
which (\ref{3.2}) becomes (see Appendix \ref{apA})
\begin{equation}
\ln g_{OCP}^{\ast }(r^{\ast })=-\Gamma r^{\ast -1}+\int d\mathbf{r}^{\ast
\prime }\left\{ g_{OCP}^{\ast }(r^{\ast \prime })-1\right\} c_{OCP}\left(
\left\vert \mathbf{r}^{\ast }-\mathbf{r}^{\ast \prime }\right\vert ;\Gamma
\right) -\Gamma B_{OCP}\left( r^{\ast }\mid g_{OCP}^{\ast }\right) ,
\label{3.2b}
\end{equation}%
together with the Ornstein-Zernicke equation for the direct correlation
function $c_{OCP}$
\begin{equation}
\left( g_{OCP}^{\ast }(r^{\ast })-1\right) =c_{OCP}\left( r^{\ast }\right)
+\int d\mathbf{r}^{\ast \prime }\left\{ g_{OCP}^{\ast }(r^{\ast \prime
})-1\right\} c_{OCP}\left( \left\vert \mathbf{r}^{\ast }-\mathbf{r}^{\ast
\prime }\right\vert ;\Gamma \right) .  \label{3.2c}
\end{equation}%
The function $B_{OCP}\left( r^{\ast }\mid g_{OCP}^{\ast }\right) $ is
referred to as the OCP "bridge function" \cite{Hansen}.

A similar rearrangement of (\ref{3.2}) is possible when $V_{0}$ is chosen to
be a harmonic trap using a different representation for $F_{ex}\left( \beta
\mid n\right) $, to obtain an exact equation for the trap density $n(r)$. In
this case the relevant reference density far from the trap center is zero.
The analysis is outlined in Appendix \ref{apA} and the resulting form for
Eq. (\ref{3.2}) is
\begin{equation}
\ln \frac{n(r)\lambda ^{3}}{z}=-\Gamma \frac{1}{2}r^{\ast 2}+\int d\mathbf{r}%
^{\ast \prime }n^{\ast }(r^{\ast \prime })c_{OCP}\left( \left\vert \mathbf{r}%
^{\ast }-\mathbf{r}^{\ast \prime }\right\vert ;\Gamma \right) -\Gamma
B\left( r^{\ast }\mid n^{\ast }\right) .  \label{3.2a}
\end{equation}%
The second term is expressed in terms of the correlations for the uniform
OCP, $c_{OCP}\left( r^{\ast };\Gamma \right) $, evaluated at the average
trap density $\overline{n}$ of (\ref{2.6}). The function $B\left( r^{\ast
}\mid n^{\ast }\right) $ differs from $B_{OCP}\left( r^{\ast }\mid
g_{OCP}^{\ast }\right) $, but by analogy will be referred to as the trap
bridge function. Equations (\ref{3.2b}) - (\ref{3.2a}) provide the formally
exact description from which approximations are formulated for the trap
density. Further details of their context are given in the Appendix.

Correlations among charges occur in these expressions through the bridge
functions and $c_{OCP}$. The simplest approximation is the "mean field
theory" resulting from the neglect of all correlations, $B\rightarrow
0,B_{OCP}\rightarrow 0,$ and $c_{OCP}\left( r^{\ast }\right) \rightarrow
-\Gamma r^{\ast -1}$, for which (\ref{3.2a}) and (\ref{3.2c}) become
\begin{equation}
\Gamma ^{-1}\ln \frac{n(r)\lambda ^{3}}{z}=-\frac{1}{2}r^{\ast 2}-\int d%
\mathbf{r}^{\ast \prime }n^{\ast }(r^{\ast \prime })\left\vert \mathbf{r}%
^{\ast }-\mathbf{r}^{\ast \prime }\right\vert ^{-1},  \label{3.3}
\end{equation}%
\begin{equation}
\Gamma ^{-1}\left( g_{MF}^{\ast }(r^{\ast })-1\right) =-r^{\ast -1}-\int d%
\mathbf{r}^{\ast \prime }\left\{ g_{MF}^{\ast }(r^{\ast \prime })-1\right\}
\left\vert \mathbf{r}^{\ast }-\mathbf{r}^{\ast \prime }\right\vert ^{-1}.
\label{3.4}
\end{equation}%
Equation (\ref{3.4}) gives the Debye-H\"{u}ckel result for a weakly coupled
OCP. The solution to (\ref{3.3}) for the mean field description of the trap
density is discussed below. This mean field equation was studied for Coulomb
and Yukawa traps in references \cite{Christian2006,Christian2007} for the
special case $\Gamma ^{-1}=T^{\ast }=0$.

To account for correlations the HNC approximation is used, which results
from neglecting the bridge functions $B\rightarrow 0,B_{OCP}\rightarrow 0$
but retaining the effects of $c_{OCP}$ (which in this
approximation is denoted $c_{HNC}$)
\begin{equation}
\ln \frac{n(r)\lambda ^{3}}{z}=-\Gamma \frac{1}{2}r^{\ast 2}+\int d\mathbf{r}%
^{\ast \prime }n^{\ast }(r^{\ast \prime })c_{HNC}\left( \left\vert \mathbf{r}%
^{\ast }-\mathbf{r}^{\ast \prime }\right\vert ;\Gamma \right) ,  \label{3.5}
\end{equation}%
\begin{equation}
\ln g_{HNC}^{\ast }(r^{\ast })=-\Gamma r^{\ast -1}+\int d\mathbf{r}^{\ast
\prime }\left\{ g_{HNC}^{\ast }(r^{\ast \prime })-1\right\} c_{HNC}\left(
\left\vert \mathbf{r}^{\ast }-\mathbf{r}^{\ast \prime }\right\vert ;\Gamma
\right) ,  \label{3.6}
\end{equation}%
\begin{equation}
\left( g_{HNC}^{\ast }(r^{\ast })-1\right) =c_{HNC}\left( r^{\ast }\right)
+\int d\mathbf{r}^{\ast \prime }\left\{ g_{HNC}^{\ast }(r^{\ast \prime
})-1\right\} c_{HNC}\left( \left\vert \mathbf{r}^{\ast }-\mathbf{r}^{\ast
\prime }\right\vert ;\Gamma \right) .  \label{3.7}
\end{equation}%
Equations (\ref{3.6}) and (\ref{3.7}) are a closed set of equations to
determine $g_{HNC}^{\ast }$ and $c_{HNC}$ for the OCP, a well-studied
problem \cite{Hansen}. Equation (\ref{3.5}) is an extension of the HNC to an
approximation for the localized trap density. It is shown below that the
inclusion of correlations leads to qualitatively different results for the
density compared to results from the mean field theory.

To go beyond the HNC approximation requires some estimate of the bridge
functions. This is a difficult problem in general. For the OCP, effects of $%
B_{OCP}$ are important only at very strong coupling. However, for the trap
density the bridge functions have a quantitative importance even at moderate
coupling. This is discussed in Section \ref{sec6} where a phenomenological
estimate is explored and shown to remove the observed discrepancies between
HNC and Monte Carlo simulation results. Further work in this direction is
planned for future studies.

In the following it is convenient to write the above equations using a
representation for the density in terms of a dimensionless effective
potential $U^{\ast }\left( r^{\ast }\right) $, defined by
\begin{equation}
n^{\ast }(r^{\ast })\equiv \overline{N}\frac{e^{-\Gamma U^{\ast }\left(
r^{\ast }\right) }}{4\pi \int_{0}^{R^{\ast }}dr^{\ast }r^{\ast 2}e^{-\Gamma
U^{\ast }\left( r^{\ast }\right) }}.  \label{3.8}
\end{equation}%
Here the chemical potential $\mu $ has been eliminated in favor of the
average number of particles $\overline{N}=\int d\mathbf{r}n(r)$.
Substitution of (\ref{3.8}) into (\ref{3.2a}) gives the corresponding exact
equation for $U^{\ast }\left( r^{\ast }\right) $%
\begin{equation}
U^{\ast }\left( r^{\ast },\Gamma ,\overline{N}\right) =\frac{1}{2}r^{\ast 2}+%
\overline{N}\frac{\int d\mathbf{r}^{\ast \prime }e^{-\Gamma U^{\ast }\left(
r^{\ast \prime },\Gamma ,\overline{N}\right) }\overline{c}(|\mathbf{r}^{\ast
}-\mathbf{r}^{\ast \prime }|,\Gamma )}{\int d\mathbf{r}^{\ast \prime
}e^{-\Gamma U^{\ast }\left( r^{\ast \prime }\Gamma ,\overline{N}\right) }}%
+B\left( r\mid n\right) ,  \label{3.9}
\end{equation}%
with the notation $\overline{c}(r^{\ast },\Gamma )\equiv -\Gamma c(r^{\ast
},\Gamma )$. The solution is parameterized by only two dimensionless
constants, $\Gamma ,\overline{N}$.

\section{Mean Field Theory}

\label{sec4}

To develop a qualitative understanding of the dependence of the profile on
the parameters $\Gamma $ and $\overline{N}$ it is useful to explore this
simple mean field approximation in the dimensionless variables of the last
section. Taking the mean field limit of (\ref{3.9}), and performing the
angular integration in the second term gives the mean field equation
\begin{eqnarray}
U^{\ast }\left( r^{\ast },\Gamma ,\overline{N}\right) &=&\frac{1}{2}r^{\ast
2}+\frac{\overline{N}}{\int_{0}^{R^{\ast }}dr^{\ast }r^{\ast 2}e^{-\Gamma
U^{\ast }\left( r^{\ast },\Gamma ,\overline{N}\right) }}  \notag \\
&&\times \left( \frac{1}{r^{\ast }}\int\limits_{0}^{r^{\ast }}dr^{\ast
\prime }\ r^{\ast \prime 2}e^{-\Gamma U^{\ast }\left( r^{\ast },\Gamma ,%
\overline{N}\right) }+\int\limits_{r^{\ast }}^{R^{\ast }}dr^{\ast \prime }\
r^{\ast \prime }e^{-\Gamma U^{\ast }\left( r^{\ast },\Gamma ,\overline{N}%
\right) }\right) .  \label{3.10}
\end{eqnarray}%
For high temperatures $T^{\ast }=\Gamma ^{-1}>>1$ (\ref{3.10}) gives%
\begin{equation}
U^{\ast }\left( r^{\ast },\Gamma ,\overline{N}\right) \rightarrow \frac{1}{2}%
\left( 1-\frac{\overline{N}}{R^{\ast 3}}\right) r^{\ast 2}+\frac{3\overline{N%
}}{2R^{\ast }},  \label{3.11}
\end{equation}%
where an outer hard wall at $R^{\ast }$ has been restored. Consequently the
trap density becomes
\begin{equation}
n^{\ast }\left( r^{\ast },\Gamma ,\overline{N}\right) \longrightarrow \frac{%
\overline{N}}{4\pi \int_{0}^{R^{\ast }}dr^{^{\prime }}r^{^{\prime
}2}e^{-\Gamma \frac{1}{2}\left( 1-\frac{\overline{N}}{R^{\ast 3}}\right)
r^{^{\prime }2}}}e^{-\Gamma \frac{1}{2}\left( 1-\frac{\overline{N}}{R^{\ast
3}}\right) r^{\ast 2}}.  \label{3.12}
\end{equation}%
It is understood that $r^{\ast }\leq R^{\ast }$ and the density is zero
otherwise. For large $R^{\ast }$ (\ref{3.12}) is just the expected Boltzmann
factor for uncorrelated particles in an external harmonic potential.
However, as the number of particles in the trap increases their Coulomb
repulsion competes with the confining trap potential, enhancing the density
toward the outer wall. For $\overline{N}<R^{\ast 3}$ the density increases
toward the center of the trap, while for $\overline{N}>R^{\ast 3}$ it
increases in the direction of the outer wall. The density is uniform at the
special \ value $R^{\ast }=\overline{N}^{1/3}=R_{0}^{\ast }$, the point
where the Coulomb repulsive force and attractive confinement force exactly
balance.
\begin{figure}[tbp]
\includegraphics{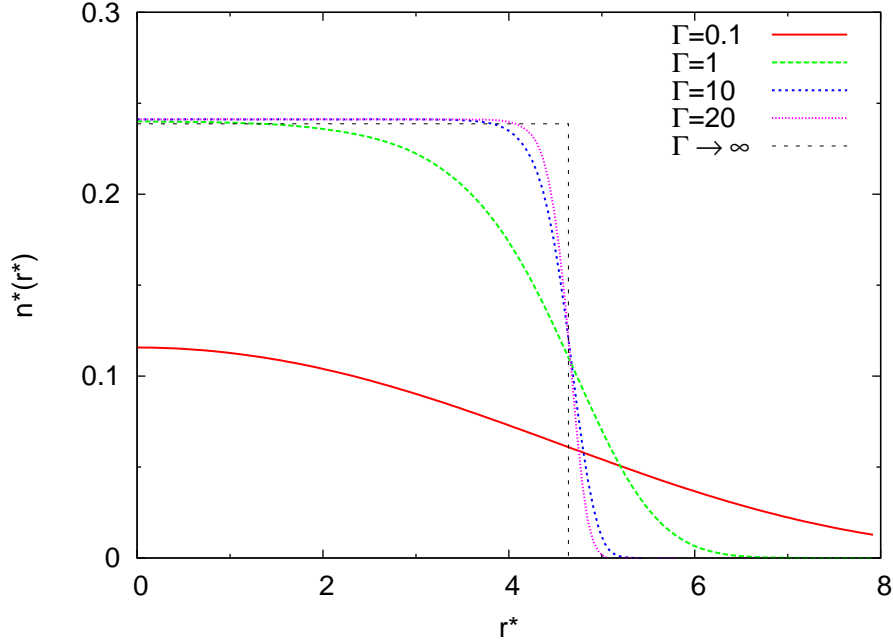}
\caption{(color online) Mean field results for $\overline{N}=100$ for
various values of $\Gamma $. The Coulomb limit ($\Gamma \rightarrow \infty $%
) is also shown.}
\label{fig1}
\end{figure}

At low temperatures $T^{\ast }=\Gamma ^{-1}<<1$ and $e^{-\Gamma U^{\ast
}\left( r^{\ast }\right) }\rightarrow 0$ in (\ref{3.10}) unless $U^{\ast
}\left( r^{\ast }\right) $ is constant, in which case these factors cancel
in the numerator and denominator. Therefore, a solution is sought of the
form
\begin{equation}
U^{\ast }\left( r^{\ast },\Gamma ,\overline{N}\right) =\left\{
\begin{array}{c}
U_{0}^{\ast },\hspace{0.25in}r^{\ast }<r_{\max }^{\ast } \\
U^{\ast }\left( r^{\ast }\right) >0,\hspace{0.25in}r^{\ast }>r_{\max }^{\ast
}%
\end{array}%
\right. ,  \label{3.13}
\end{equation}%
for some constant $U_{0}^{\ast }$ and $r_{\max }^{\ast }<R^{\ast }$. Then,
for $\Gamma >>1$ (\ref{3.10}) gives for $r^{\ast }<r_{\max }^{\ast }$
\begin{equation}
U^{\ast }\left( r^{\ast },\Gamma ,\overline{N}\right) \rightarrow \frac{1}{2}%
\left( 1-\frac{\overline{N}}{r_{\max }^{\ast 3}}\right) r^{\ast 2}+\frac{3%
\overline{N}}{2r_{\max }^{\ast }}  \label{3.14}
\end{equation}%
The first term must vanish for consistency with (\ref{3.13}), which
determines $r_{\max }^{\ast }$
\begin{equation}
r_{\max }^{\ast }=R_{0}^{\ast }=\overline{N}^{1/3}.  \label{3.15}
\end{equation}%
Next, for $\Gamma >>1$ and $R_{0}^{\ast }<r^{\ast }\leq R^{\ast }$ (\ref%
{3.10}) gives%
\begin{equation}
U^{\ast }\left( r^{\ast },\Gamma ,\overline{N}\right) \rightarrow \frac{1}{2}%
r^{\ast 2}+\frac{\overline{N}}{r^{\ast }}.  \label{3.16}
\end{equation}%
Consequently, the trap density in this limit is%
\begin{equation}
n^{\ast }(r^{\ast })\rightarrow \frac{3\overline{N}}{4\pi R_{0}^{\ast 3}}%
\left\{
\begin{array}{c}
1,\hspace{0.25in}r^{\ast }<R_{0}^{\ast } \\
e^{-\Gamma \left( \frac{1}{2}r^{\ast 2}+\frac{\overline{N}}{r^{\ast }}%
\right) },\hspace{0.25in}r^{\ast }>R_{0}^{\ast }%
\end{array}%
\right. .  \label{3.18}
\end{equation}%
The density is uniform up to $R_{0}^{\ast }$ which depends only on $%
\overline{N}$, and is exponentially small for larger $r^{\ast }$; the
density effectively has a step profile.

More generally, (\ref{3.10}) can be solved numerically for the full range of
$\Gamma $ and $\overline{N}$. The dependence on $\Gamma $ is illustrated in
Figure 1 for $\overline{N}=100$ and $\Gamma =0.1,1,10,$ and $20$. Also shown
is the limiting step function for $\Gamma \rightarrow \infty $. The edge of
the profile is the radius at which the confinement force equals the total
Coulomb force, $R_{0}^{\ast }=\overline{N}^{1/3}$. Although instructive as a
reference, the low temperature behavior within the mean field approximation
is relevant only for the average density, since charge correlation effects
become dominant for $\Gamma \gtrsim 10$, as shown in the next section.

\section{HNC Approximation}

\label{sec5}

The density in the HNC approximation is given by (\ref{3.5}) - (\ref{3.7}).
The effective potential is determined from the equation
\begin{equation}
U^{\ast }\left( r^{\ast },\Gamma ,\overline{N}\right) =\frac{1}{2}r^{\ast 2}+%
\overline{N}\frac{\int d\mathbf{r}^{\ast \prime }e^{-\Gamma U^{\ast }\left(
r^{\ast \prime },\Gamma ,\overline{N}\right) }\overline{c}_{HNC}(|\mathbf{r}%
^{\ast }-\mathbf{r}^{\ast \prime }|,\Gamma )}{\int d\mathbf{r}^{\ast \prime
}e^{-\Gamma U^{\ast }\left( r^{\ast \prime }\Gamma ,\overline{N}\right) }},
\label{4.1}
\end{equation}%
The only difference from the mean field form is the replacement of the
Coulomb interaction by the direct correlation function $\overline{c}_{HNC}$.
This is first calculated from the corresponding HNC approximation for the
OCP, (\ref{3.6}) and (\ref{3.7}), and the results are illustrated in Figure
2 for $\Gamma =0.1,1,10,$ and $100$. It is seen that the Coulomb limit is
recovered for all $r^{\ast }$ in the limit $\Gamma \rightarrow 0$, and for
any $\Gamma $ at sufficiently large $r^{\ast }$.
\begin{figure}[tbp]
\includegraphics{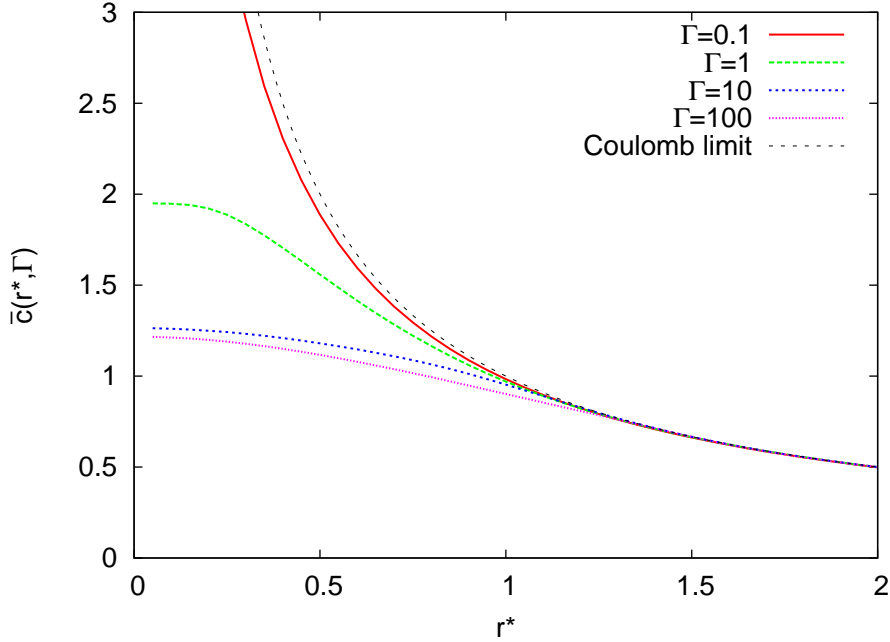}
\caption{(color online) Dependence of the scaled direct correlation function
on $\Gamma $. Also shown is the small $\Gamma $ Coulomb limit (mean-field
approximation ).}
\label{fig2}
\end{figure}

\begin{figure}[tbp]
\includegraphics{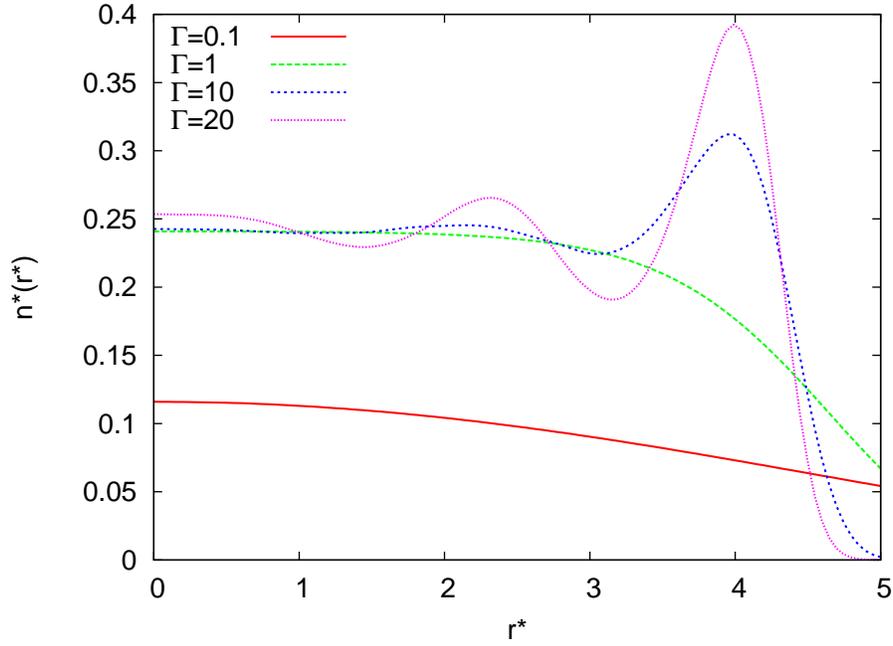}
\caption{(color online) Radial density profile with correlations included,
for the same conditions as Figure 1.}
\label{fig3}
\end{figure}

\begin{figure}[tbp]
\includegraphics{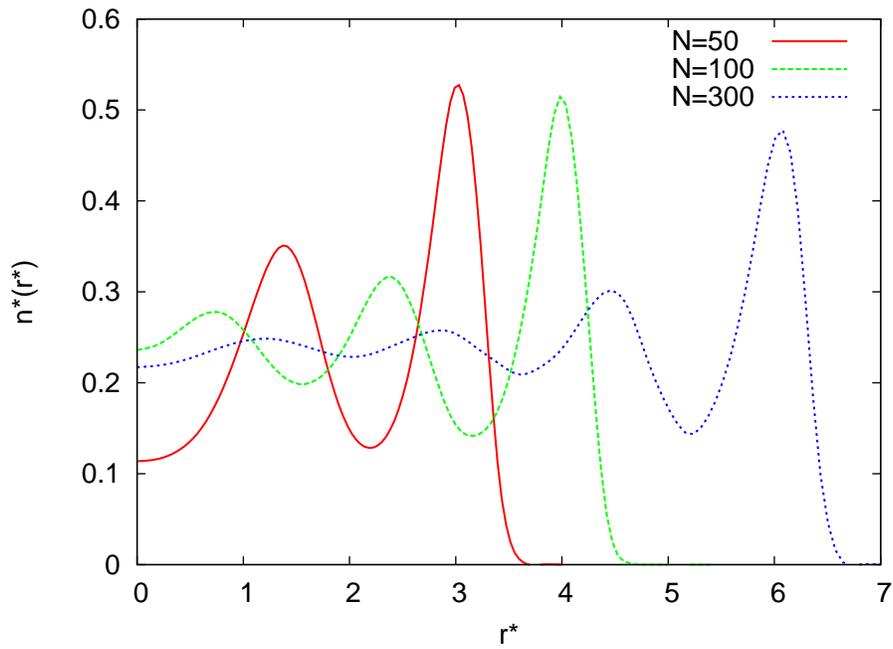}
\caption{(color online) Dependence of radial density profile on $\bar{N}$
for $\Gamma =40$.}
\label{fig4}
\end{figure}

With $\overline{c}_{HNC}$ known, (\ref{4.1}) can be solved to determine the
trap density. Figure 3 shows the results for the same conditions as in
Figure 1, illustrating the effects of correlations relative to the mean
field results. The monotonic decrease with $r^{\ast }$ is now replaced by
the appearance of structure ("shells") for $\Gamma =10$ and $20$. The latter
two curves also suggest that the location of the shells is
insensitive to the value of $\Gamma $, as is confirmed below. Figure 4, for $%
\Gamma =40$, shows that the number of "shells" increases with $\overline{N}$%
. Figure 5 illustrates the formation and filling of the first three shells,
giving the number of particles in each shell (defined as the number between
two spheres at successive radial minima) as a function of $\overline{N}$ for
$\Gamma =10$, $20$, and $40$. The initial dependence is linear as only a
single outer shell is populated. However, after formation of the second
shell the dependence is more complex as additional particles can go into
either shell. Remarkably, this process of filling is almost independent of $%
\Gamma $ (in agreement with experiment and simulations \cite%
{Baumgartner2007,Golubnychiy2006}).

\begin{figure}[tbp]
\includegraphics{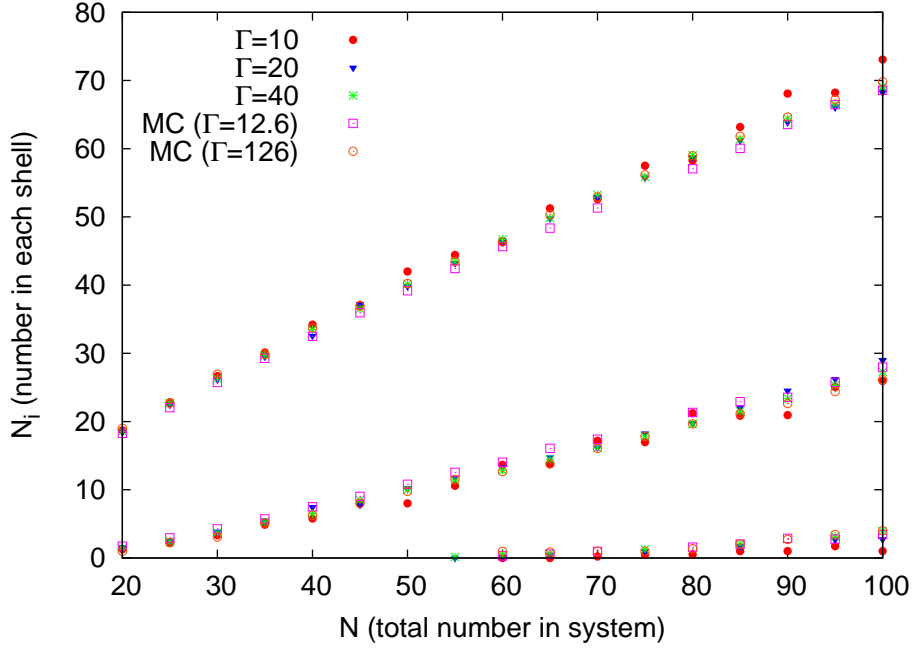}
\caption{(color online) Shell population for two-shell conditions. The
higher trend curve is for the the outer shell. Data is shown for HNC and MC calculations.} 
\label{fig5}
\end{figure}

\section{Comparison to Monte Carlo Simulation}

\label{sec6}To assess the quality of predictions based on the HNC
approximation some benchmark Monte Carlo simulations have been performed for
comparison. For moderate coupling, $\Gamma =1$, correlations are important
(see Figure 2, for example) but shell structure does not appear. Figure 6
shows the comparison with Monte Carlo results for $\Gamma =1$, and $%
\overline{N}=10,100,$ and $1000$. Not shown are results for $\overline{N}<10$
where the agreement is poor even at weak coupling. The first conclusion is
that the HNC approximation is accurate at weak to moderate coupling, $\Gamma
\leq 1$, and $\overline{N}\geq 10$.

\begin{figure}[tbp]
\includegraphics{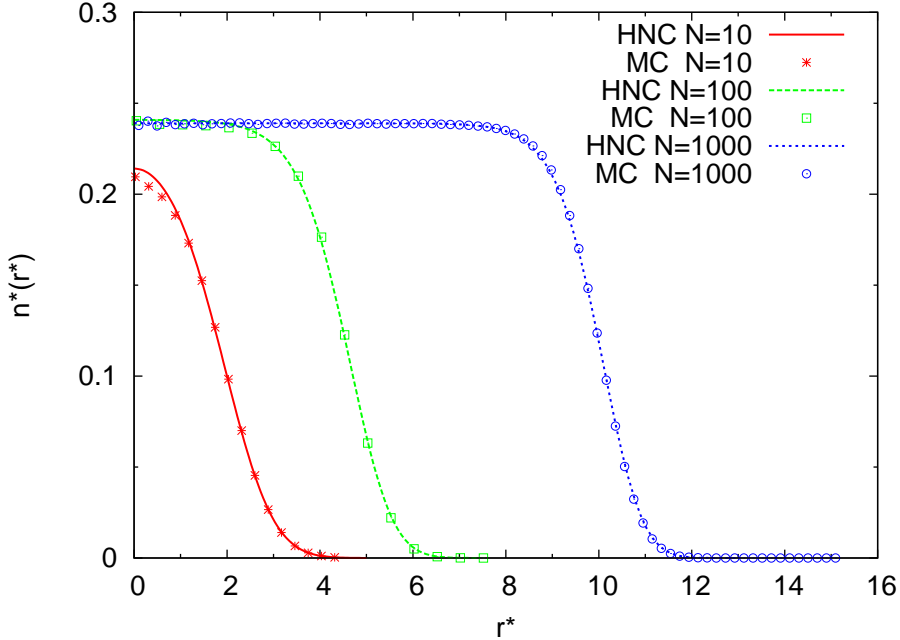}
\caption{(color online) Comparison with Monte Carlo results for $\Gamma=1$.}
\label{fig6}
\end{figure}

\begin{figure}[tbp]
\includegraphics{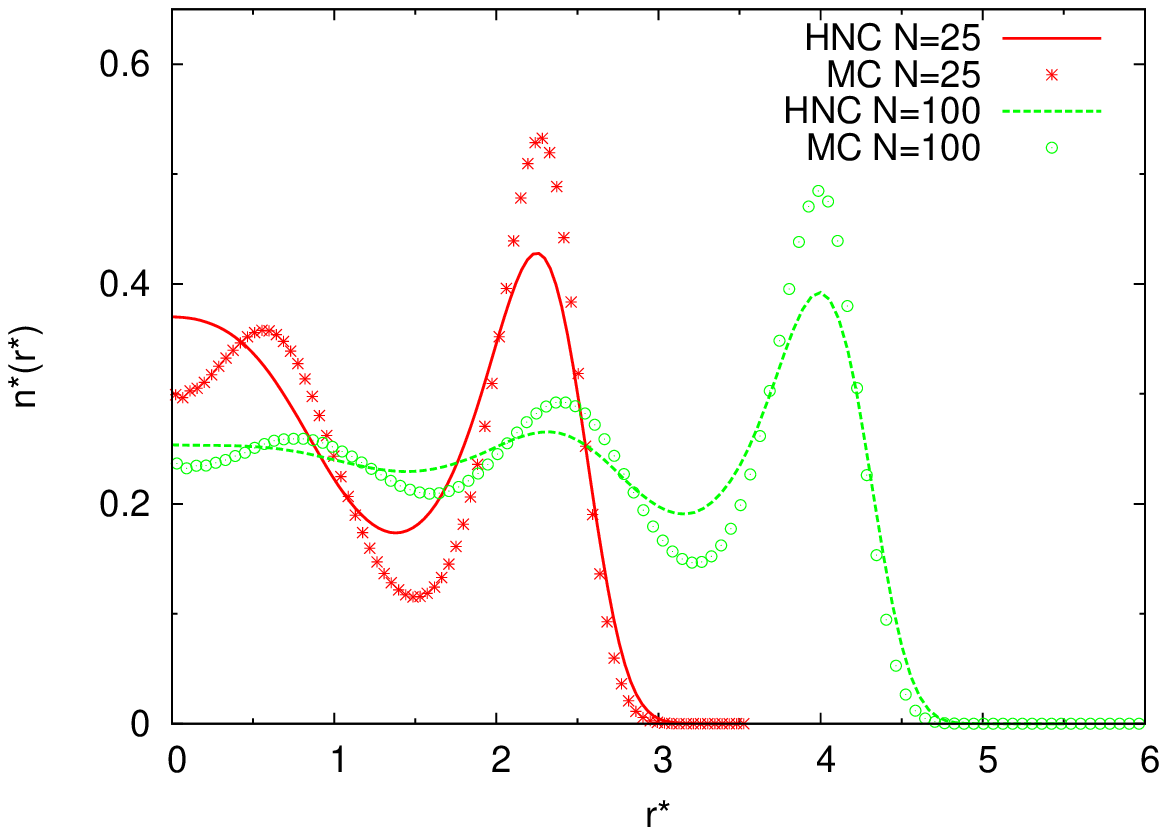}
\caption{(color online) Comparison with Monte Carlo results for $\Gamma=20$.}
\label{fig7}
\end{figure}


\begin{figure}[tbp]
\includegraphics{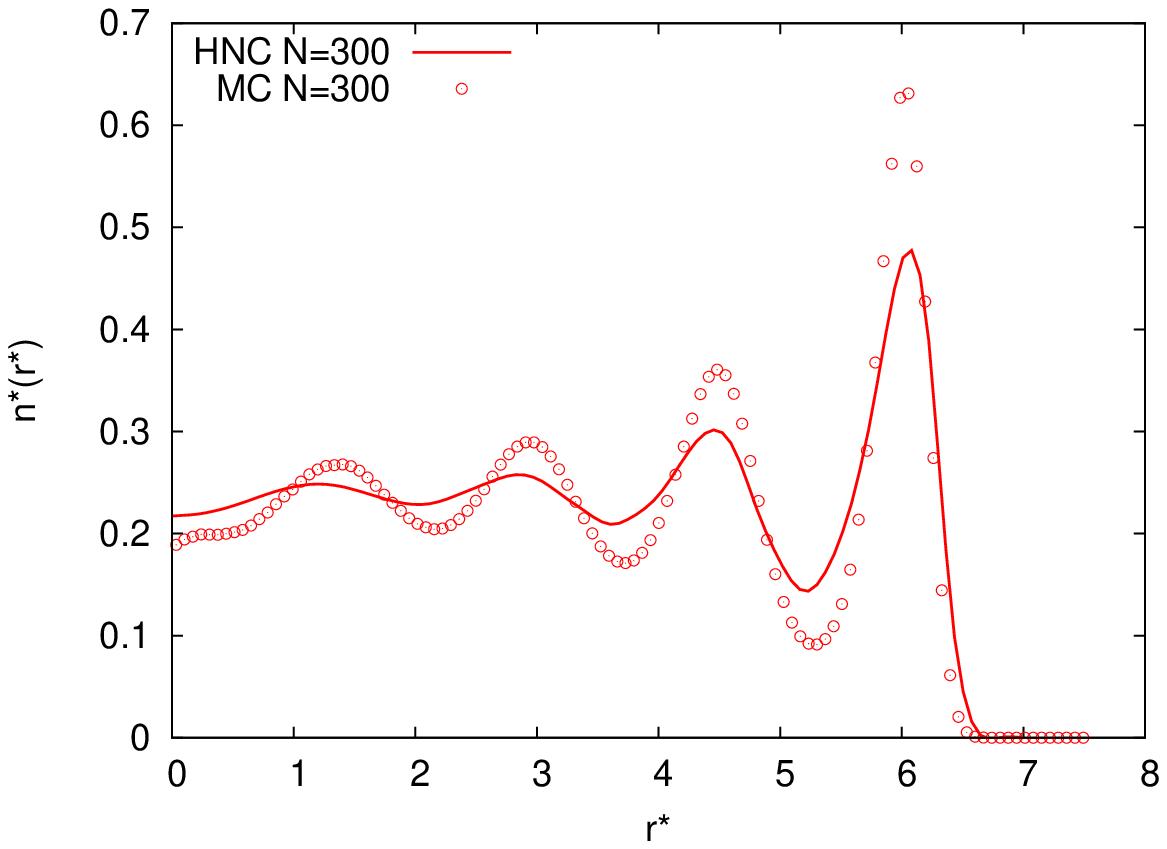}
\caption{(color online) Comparison with Monte Carlo results for $\bar{N}=300$
and $\Gamma =40$.}
\label{fig8}
\end{figure}

The formation of shell structure at larger values of $\Gamma $ is tested in
Figure 7 for $\Gamma =20,$ $\overline{N}=25$ and $100$, and in Figure 8 for $%
\Gamma =40,$ $\overline{N}=300$. It is seen that the formation and location
of the peaks is very well described, but systematically their widths are too
large and amplitudes too small by as much as $20-40\%$. Nevertheless, the
particle number in the shells is quite accurate, as shown in Figure \ref%
{fig5}.
\begin{figure}[tbp]
\includegraphics{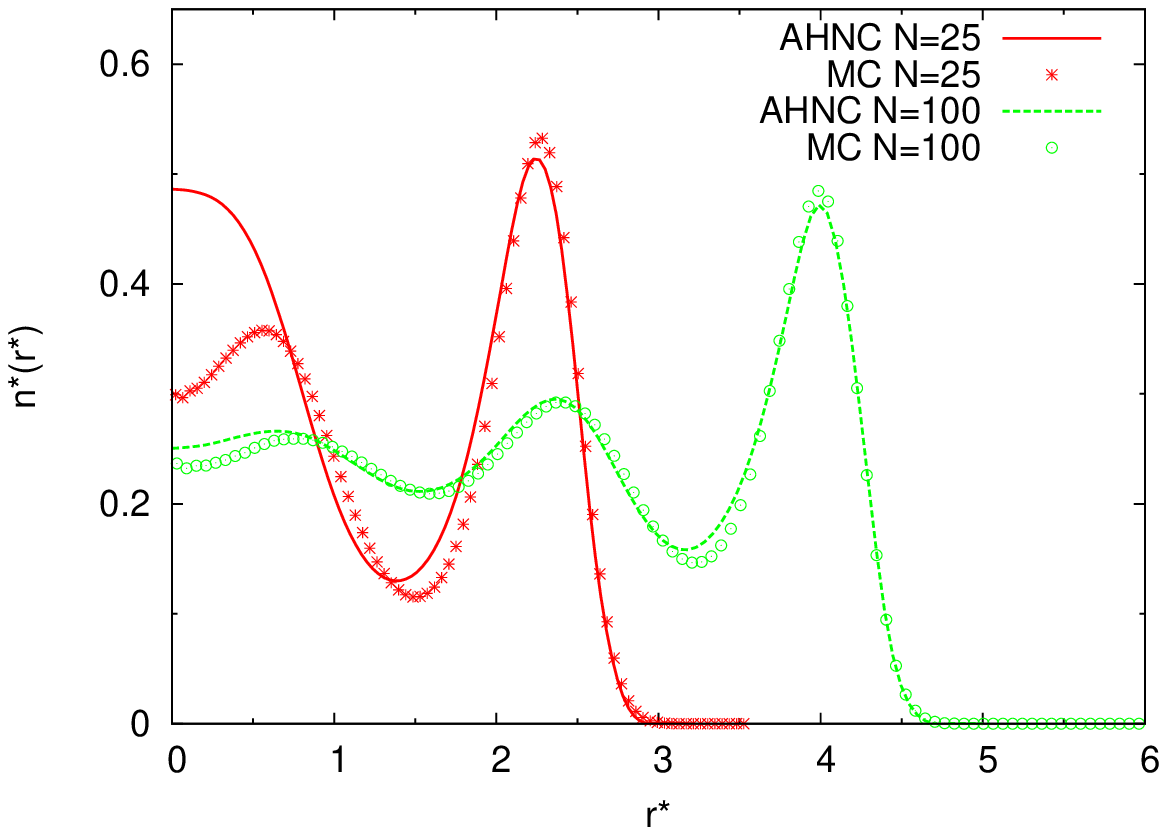}
\caption{(color online) Comparison of adjusted HNC with Monte Carlo results
for $\Gamma =20$. }
\label{fig9}
\end{figure}

\begin{figure}[tbp]
\includegraphics{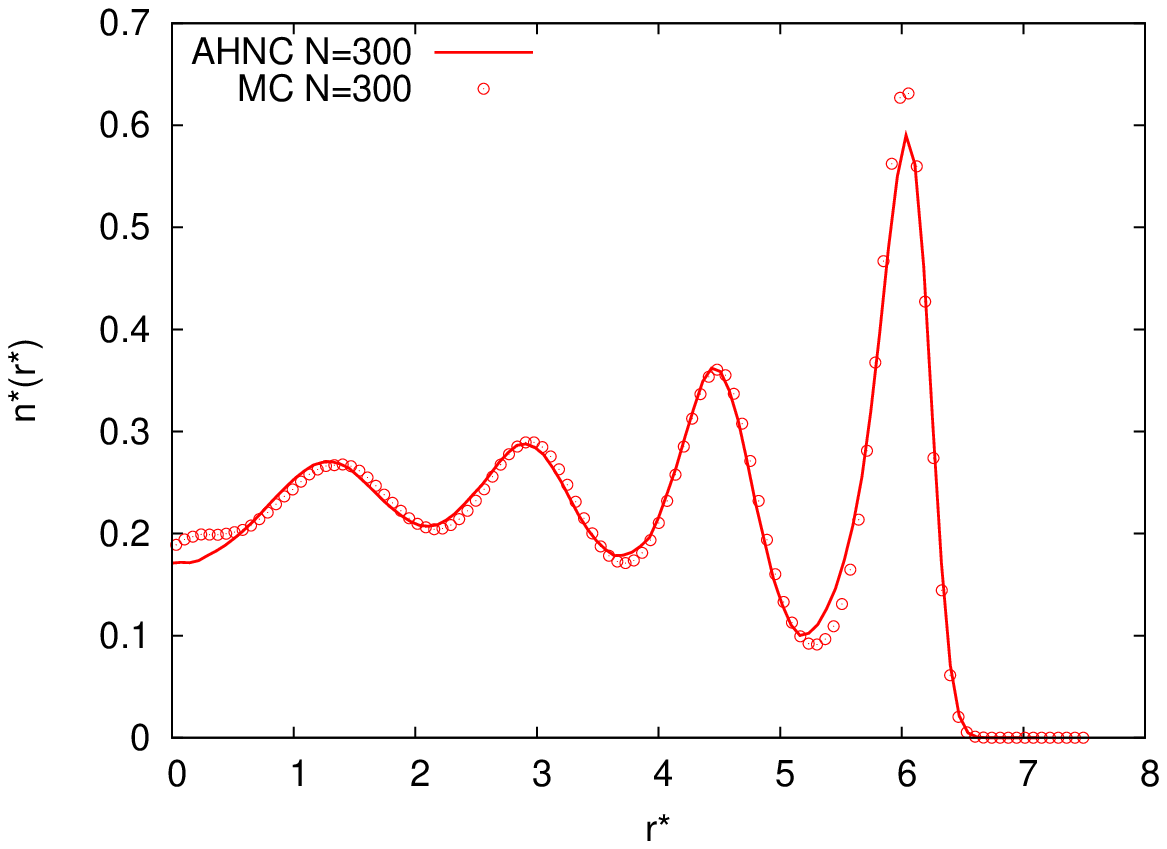}
\caption{(color online) Comparison of adjusted HNC with Monte Carlo results
for $\bar{N}=300 $ and $\Gamma =40$.}
\label{fig10}
\end{figure}

\section{Adjusted HNC}

\label{sec7}The HNC approximation provides a surprisingly good
representation of correlation effects on the density profile of strongly
coupled charges in a trap: formation, population, location, and number of
shells. Its primary limitation is the significant discrepancy in the
amplitudes and widths of the shells. This is due to neglect of important
contributions from the bridge functions $B\left( r\mid n\right) $ and $%
B_{OCP}\left( r\mid n\right) $ in (\ref{3.2b}) and (\ref{3.2a}) at strong
coupling. This problem was addressed some time ago for the OCP by Ng \cite%
{Ng}, who observed that the peaks of the HNC pair distribution function $%
g_{OCP}^{\ast }(r^{\ast })$ appear at the correct position but with \
incorrect amplitudes and widths. This is clearly analogous to the current
errors of the trap density. Ng corrected the HNC by an estimate of the OCP
bridge function in the form%
\begin{equation}
B_{OCP}\left( r\mid n\right) \rightarrow \lambda \left( \Gamma \right)
V_{0}\left( r^{\ast }\right) ,\hspace{0.25in}V_{0}\left( r^{\ast }\right) =%
\frac{1}{r^{\ast }},  \label{6.1}
\end{equation}%
where $\lambda \left( \Gamma \right) $ is a chosen function to interpolate
between $\lambda \left( 0\right) =0$ and some constant value $\lambda \left(
\infty \right) $. The advantage of the chosen form is that when inserted in (%
\ref{3.2b}) the form of the HNC approximation is recovered, but with a
renormalized coupling constant $\Gamma ^{\prime }=\left( 1+\lambda \left(
\Gamma \right) \right) \Gamma $
\begin{equation}
\ln g_{OCP}^{\ast }(r^{\ast })=-\Gamma ^{\prime }r^{\ast -1}+\int d\mathbf{r}%
^{\ast \prime }\left( g_{OCP}^{\ast }(r^{\ast \prime })-1\right)
c_{OCP}\left( \left\vert \mathbf{r}^{\ast }-\mathbf{r}^{\ast \prime
}\right\vert ;\Gamma ^{\prime }\right) .  \label{6.2}
\end{equation}%
Note that $c_{OCP}\left( \left\vert \mathbf{r}^{\ast }-\mathbf{r}^{\ast
\prime }\right\vert ;\Gamma ^{\prime }\right) $ also becomes a function of
the renormalized coupling constant since it inherits that dependence only
from $g_{OCP}^{\ast }(r^{\ast })$ via the Ornstein-Zernicke equation (\ref%
{3.2c}). Ng found that the Monte Carlo data for $g_{OCP}^{\ast }(r^{\ast })$
could be fit within two percent with the choice $\lambda \left( \infty
\right) =0.6$. Alternatively, (\ref{6.1}) can be viewed as a representation
of the OCP bridge function determined from (\ref{3.2c}) using $g_{OCP}^{\ast
}(r^{\ast })$ from Monte Carlo as input.

It is plausible to use this same scheme for the trap bridge function $%
B\left( r\mid n\right) $ as well \
\begin{equation}
B\left( r\mid n\right) \rightarrow \lambda \left( \Gamma \right) V_{0}\left(
r^{\ast }\right) ,\hspace{0.25in}V_{0}\left( r^{\ast }\right) =\frac{1}{2}%
r^{\ast 2},  \label{6.3}
\end{equation}%
with the same choice for $\lambda \left( \Gamma \right) $. Accordingly, (\ref%
{3.2a}) gives the HNC form for the trap density%
\begin{equation}
\ln \frac{n(r)\lambda ^{3}}{z}=-\Gamma ^{\prime }\frac{1}{2}r^{\ast 2}+\int d%
\mathbf{r}^{\ast \prime }n^{\ast }(r^{\ast \prime })c_{OCP}\left( \left\vert
\mathbf{r}^{\ast }-\mathbf{r}^{\ast \prime }\right\vert ;\Gamma ^{\prime
}\right) .  \label{6.4}
\end{equation}%
This is the same as the HNC approximation of Section \ref{sec5}, except with
the bridge function corrected value $\Gamma ^{\prime }$. Based on the
results of Ng, the same value of $\lambda \left( \infty \right) =0.6$ for
the strong coupling conditions is used here. Equations (\ref{6.2}) and (\ref%
{6.4}) will be referred to as the adjusted HNC (AHNC) approximation. Figures
9 and 10 show a revised comparison for the conditions of figures 7 and 8
but now with the AHNC calculated with $\Gamma ^{\prime }=1.6\Gamma $. Most
of the discrepancies are seen to be removed by this AHNC approximation. It
describes accurately the complexities of the trap density over a range of $%
\Gamma ,N$ \ including strong coupling, with the structural simplicity of
the HNC approximation and only OCP correlations as input. Its domain of
applicability and limitations will be explored in more detail elsewhere. It
is somewhat remarkable that the simple forms (\ref{6.1}) and (\ref{6.3}) for
the bridge functions work so well. This success should motivate future work
to address the theoretical basis for such practical approximations to the
bridge functions.

\section{Discussion}

\label{sec8}

The objective here has been the development and exploration of a theoretical
framework for analysis of the density profile of charges in a harmonic trap.
Previous extensive work on this problem has focused on experimental and
simulation studies of the low temperature crystal structure (and its
melting) for Coulomb, Yukawa, and Lennard-Jones interactions. Here the
complementary theoretical study starting from the high temperature fluid
phase has been described, showing the evolution of shell structure at lower
temperatures due to correlations that are precursors of the observed
structure in the crystal phase. A first order representation of correlations
in the HNC approximation is shown to capture the radial structural features
of the density profile. Particularly remarkable is the formation of
"blurred" shells for moderate coupling at the magic numbers for tiling the
sharp crystal shells. This anticipation in the fluid phase of the low
temperature crystal structure will be discussed
elsewhere.

The analysis above provides justification for a simple and practical
description within the fluid phase for charged particle confinement, the
AHNC approximation. Although the discussion here has been limited to the
case of Coulomb interactions there is no complication involved in extending
it to other more practical forms such as Yukawa particles. The formulation
of this approximation within the context of density functional theory also
provides a complete thermodynamics for this system. In this way it is hoped
that the AHNC could be improved to include a description of freezing and
crystal structure.

\section{Acknowledgements}

This work is supported by the Deutsche Forschungsgemeinschaft via SFB-TR 24,
and by the NSF/DOE Partnership in Basic Plasma Science and Engineering under
the Department of Energy award DE-FG02-07ER54946.

\appendix

\section{Representations from Density Functional Theory}

\label{apA}

Consider a given system in an external potential. For simplicity
only the case of spherically symmetric central potentials are considered.
Two classes of external potential are identified. In the first class, the
potential goes to zero or a constant at large $r,$ resulting in an average
density that becomes uniform at large distances. This is the usual case of
localized perturbations in a bulk fluid. The second class is potentials that
become unbounded for large $r$, so that the resulting average density is
localized in some region about the origin. This is the case of interest here
for charges in a trap. In the following, different exact representations of
the inhomogeneous density appropriate for each of these cases are given.

The approach is based on the formal structure of density functional theory
\cite{Evans,Lutsko}. To review briefly the relevant equations, the grand
potential is defined first by

\begin{equation}
\beta \Omega \left( \beta \mid \mu -V_{0}\right) =-\ln \Xi
\sum_{N=0}^{\infty }Tre^{-\beta \left( H-\mu N\right) },  \label{a.1}
\end{equation}%
where $\Xi $ is the grand partition function of (\ref{3.1a}). The notation
makes explicit the functional dependence of $\Omega $ on the external
potential $V_{0}$. The density is obtained from the functional derivative of
$\Omega $%
\begin{equation}
n(r)=\frac{\delta \Omega }{\delta V_{0}(r)}.  \label{a.2}
\end{equation}%
The roles of $n(r)$ and $V_{0}(r)$ can be exchanged by the Legendre
transformation
\begin{equation}
F(\beta \mid n)\equiv \Omega \left( \beta \mid \mu -V_{0}\right) +\int d%
\mathbf{r}\left( \mu -V_{0}\left( r\right) \right) n\left( r\right) ,
\label{a.3}
\end{equation}%
\begin{equation}
\frac{\delta F\left( \beta \mid n\right) }{\delta n\left( r\right) }=\mu
-V_{0}\left( r\right) .  \label{a.4}
\end{equation}%
The new functional $F\left( \beta \mid n\right) $ is the free energy
expressed as a universal functional of the density. Equations (\ref{a.2})
and (\ref{a.4}) \ make explicit the fact that the density is a functional of
the applied potential and vice versa. The functional $F\left( \beta \mid
n\right) $ is universal in the sense that each choice for $V_{0}$
corresponds to evaluation of this functional at a corresponding specific
density field. Equation (\ref{a.4}) will be applied here as an exact
equation for the density in terms of the chosen $V_{0}$.

The free energy can be divided into an "ideal gas" form $F_{0}(\beta \mid n)$%
, (its form in the absence of interactions among the particles), plus the
"excess free energy" due to interactions, $F_{ex}(\beta \mid n)$%
\begin{equation}
F(\beta \mid n)=F_{0}(\beta \mid n)+F_{ex}(\beta \mid n).  \label{a.5}
\end{equation}%
The first term can be calculated directly so that (\ref{a.4}) takes the more
practical form of (\ref{3.2})
\begin{equation}
\ln \frac{n(r)\lambda ^{3}}{z}=-\beta V_{0}\left( r\right) -\frac{\delta
\beta F_{ex}\left( \beta \mid n\right) }{\delta n\left( r\right) }.
\label{a.6}
\end{equation}

A further decomposition of $F_{ex}$ is useful for the introduction of
approximations. The central idea is to exploit knowledge of correlations in
uniform systems to characterize the non-uniform density profile due to $%
V_{0} $. First, note the identity for any functional $X(g)$ of the function $%
g\left( \mathbf{r}\right) $
\begin{eqnarray}
X(g)-X(g_{r}) &=&\int_{0}^{1}d\lambda \partial _{\lambda }X\left[ \lambda
g+\left( 1-\lambda \right) g_{r}\right]  \notag \\
&=&\int_{0}^{1}d\lambda \int d\mathbf{r}\frac{\delta X\left( g\right) }{%
\delta g\left( \mathbf{r}\right) }\mid _{\lambda g+\left( 1-\lambda \right)
g_{r}}\left( g\left( \mathbf{r}\right) -g_{r}\left( \mathbf{r}\right) \right)
\label{a.7}
\end{eqnarray}%
where $g_{r}\left( \mathbf{r}\right) $ is an arbitrary reference function.
Application of this identity twice to the excess free energy $F_{ex}$ gives%
\begin{align}
F_{ex}(\beta & \mid n)=F_{ex}(\beta \mid n_{r})-\int d\mathbf{r}\beta
^{-1}c^{(1)}\left( \mathbf{r\mid }n_{r}\right) \left( n\left( \mathbf{r}%
\right) -n_{r}\right)  \notag \\
& -\int_{0}^{1}d\lambda \int_{0}^{\lambda }d\lambda ^{\prime }\int d\mathbf{r%
}d\mathbf{r}^{\prime }\beta ^{-1}c^{(2)}\left[ \mathbf{r,r}^{\prime }\mathbf{%
\mid }\lambda ^{\prime }n+\left( 1-\lambda ^{\prime }\right) n_{r}\right]
\left( n\left( \mathbf{r}\right) -n_{r}\right) \left( n\left( \mathbf{r}%
^{\prime }\right) -n_{r}\right) .  \label{a.8}
\end{align}%
Here $n_{r}$ is an arbitrary reference density, and the direct correlation
functions have been introduced by the definitions%
\begin{equation}
c^{(m)}\left( \mathbf{r}_{1},\mathbf{r}_{2}..\mathbf{r}_{m}\mid n\right)
\equiv -\beta \frac{\delta ^{m}F_{ex}\left( \beta \mid n\right) }{\delta
n\left( \mathbf{r}_{1}\right) \delta n\left( \mathbf{r}_{2}\right) ..\delta
n\left( \mathbf{r}_{m}\right) }.  \label{a.9}
\end{equation}

For external potentials $V_{0}$ that vanish for large $r$ there is a natural
reference density $\overline{n}$ corresponding to the uniform system at
distances far from the perturbation. Then (\ref{a.2}) becomes%
\begin{align}
F_{ex}(\beta & \mid n)=F_{ex}(\beta \mid \overline{n})-\int_{0}^{1}d\lambda
\left( 1-\lambda \right)  \notag \\
& \times \int d\mathbf{r}d\mathbf{r}^{\prime }\beta ^{-1}c^{(2)}\left[
\mathbf{r,r}^{\prime }\mathbf{\mid }\lambda ^{\prime }n+\left( 1-\lambda
^{\prime }\right) \overline{n}\right] \left( n\left( \mathbf{r}\right) -%
\overline{n}\right) \left( n\left( \mathbf{r}^{\prime }\right) -\overline{n}%
\right) .  \label{a.10}
\end{align}%
Here the second term of (\ref{a.8}) vanishes by requiring that $\overline{n}%
V=\overline{N}$, the average total particle number. Again applying (\ref{a.7}%
) to express $c^{(2)}\left( \mathbf{r,r}^{\prime }\mathbf{\mid }\lambda
^{\prime }n+\left( 1-\lambda ^{\prime }\right) \overline{n}\right) $ in
terms of the uniform density gives%
\begin{align}
F_{ex}(\beta & \mid n)=F_{ex}(\beta \mid \overline{n})-\frac{1}{2}\int d%
\mathbf{r}d\mathbf{r}^{\prime }\beta ^{-1}c^{(2)}\left( \left\vert \mathbf{%
r-r}^{\prime }\right\vert \mathbf{\mid }\overline{n}\right) \left( n\left(
\mathbf{r}\right) -\overline{n}\right) \left( n\left( \mathbf{r}^{\prime
}\right) -\overline{n}\right)  \notag \\
& -\int_{0}^{1}d\lambda \left( 1-\lambda \right) \int_{0}^{\lambda }d\lambda
^{\prime }\int d\mathbf{r}d\mathbf{r}^{\prime }d\mathbf{r}^{\prime \prime
}\left( n\left( \mathbf{r}\right) -\overline{n}\right) \left( n\left(
\mathbf{r}^{\prime }\right) -\overline{n}\right) \left( n\left( \mathbf{r}%
^{\prime \prime }\right) -\overline{n}\right)  \notag \\
& \times c^{(3)}\left[ \mathbf{r},\mathbf{r}^{\prime },\mathbf{r}^{\prime
\prime }\mid \lambda ^{\prime }n+(1-\lambda ^{\prime })\overline{n}\right]
\label{a.11}
\end{align}

Next, consider external potentials that become unbounded for large $r$ so
that the density vanishes at large distances. In this case the appropriate
choice for the reference density in (\ref{a.8}) is $n_{r}=0$, giving
\begin{equation}
F_{ex}(\beta \mid n)=\int_{0}^{1}d\lambda \left( 1-\lambda \right) \int d%
\mathbf{r}d\mathbf{r}^{\prime }n\left( \mathbf{r}\right) n\left( \mathbf{r}%
^{\prime }\right) \beta ^{-1}c^{(2)}\left( \mathbf{r},\mathbf{r}^{\prime
}\mid \lambda n\right) ,  \label{a.12}
\end{equation}%
The first two terms on the right side of (\ref{a.8}) vanish at zero density,
and the first $\lambda $ integral has been performed. The integration
domains are now controlled by the density, which restricts them to the
bounded domain. In this domain, it is reasonable to identify an appropriate
uniform reference density $\overline{n}$ and to express $c^{(2)}\left(
\mathbf{r},\mathbf{r}^{\prime }\mid \lambda n\right) $ in terms of this
uniform density by a further application of (\ref{a.7})
\begin{align}
c^{(2)}\left( \mathbf{r},\mathbf{r}^{\prime }\mid \lambda n\right) &
=c^{(2)}\left( \mathbf{r},\mathbf{r}^{\prime }\mid \overline{n}\right)
+\int_{0}^{1}dx\partial _{x}c^{(2)}\left[ \mathbf{r},\mathbf{r}^{\prime
}\mid x\lambda n+(1-x)\overline{n}\right]  \notag \\
& =c^{(2)}\left( \mathbf{r},\mathbf{r}^{\prime }\mid \overline{n}\right)
\notag \\
& +\int_{0}^{1}dx\int d\mathbf{r}^{\prime \prime }\left( \lambda n\left(
\mathbf{r}^{\prime \prime }\right) -\overline{n}\right) c^{(3)}\left[
\mathbf{r},\mathbf{r}^{\prime },\mathbf{r}^{\prime \prime }\mid x\lambda
n+(1-x)\overline{n}\right] .  \label{a.13}
\end{align}%
Then the excess free energy becomes%
\begin{align}
F_{ex}(\beta & \mid n)=-\frac{1}{2}\int d\mathbf{r}d\mathbf{r}^{\prime
}n\left( \mathbf{r}\right) n\left( \mathbf{r}^{\prime }\right) \beta
^{-1}c^{(2)}\left( \left\vert \mathbf{r-r}^{\prime }\right\vert \mid
\overline{n}\right)  \notag \\
& -\int_{0}^{1}d\lambda \left( 1-\lambda \right) \int_{0}^{1}dx\int d\mathbf{%
r}^{\prime \prime }\beta ^{-1}c\left[ ^{(3)}\mathbf{r},\mathbf{r}^{\prime },%
\mathbf{r}^{\prime \prime }\mid x\lambda n+(1-x)\overline{n}\right] \left(
\left( \lambda n\left( \mathbf{r}^{\prime \prime }\right) -\overline{n}%
\right) \right) ,  \label{a.14}
\end{align}

Equations (\ref{a.11}) and (\ref{a.14}) are both formally exact and
equivalent, but each is formulated so that the leading terms are reasonable
approximations for different classes of external potentials. Use of (\ref%
{a.11}) in (\ref{a.6}) leads to the form (\ref{3.2b}) of the text; use of (%
\ref{a.14}) in (\ref{a.6}) leads to the form (\ref{3.2a}). The
Ornstein-Zernicke equation (\ref{3.2c}) is an identity that follows from the
definition of $c^{(2)}$\ in (\ref{a.9}). The last terms of (\ref{a.11}) and (%
\ref{a.14}) give rise to what is referred to as bridge functions in the
text, and provide the formal definitions for each.

The above is quite general. For the case of interest here, a system of
charges, the uniform density case corresponds to the one component plasma.

\end{document}